\documentclass[runningheads]{llncs}
\usepackage{graphicx}
\usepackage{cite}
\usepackage{float}
\usepackage{booktabs}
\begin{document}

\title{Machine Learning for Network-based Intrusion Detection Systems: an Analysis of the CIDDS-001 Dataset}
\titlerunning{Machine Learning for Network-based Intrusion Detection Systems}
%

\author{José Carneiro\orcidID{0000-0003-2695-8535} \and
Nuno Oliveira\orcidID{0000-0002-5030-7751} \and \\
Norberto Sousa\orcidID{0000-0003-2919-4817} \and Eva Maia\orcidID{0000-0002-8075-531X} \and \\ Isabel Praça\orcidID{0000-0002-2519-9859}}
\authorrunning{J. Carneiro et al.}
%
\institute{Research Group on Intelligent Engineering and Computing for Advanced Innovation and Development (GECAD), Porto School of Engineering (ISEP), 4200-072 Porto, Portugal \\ 
\email{\{emcro,nunal,norbe,egm,icp\}@isep.ipp.pt}}

\maketitle              
\begin{abstract}
With the increasing amount of reliance on digital data and computer networks by corporations and the public in general, the occurrence of cyber attacks has become a great threat to the normal functioning of our society. Intrusion detection systems seek to address this threat by preemptively detecting attacks in real time while attempting to block them or minimizing their damage. These systems can function in many ways being some of them based on artificial intelligence methods. Datasets containing both normal network traffic and cyber attacks are used for training these algorithms so that they can learn the underlying patterns of network-based data. The CIDDS-001 is one of the most used datasets for network-based intrusion detection research. Regarding this dataset, in the majority of works published so far, the \textit{Class} label was used for training machine learning algorithms. However, there is another label in the CIDDS-001, \textit{AttackType}, that seems very promising for this purpose and remains considerably unexplored. This work seeks to make a comparison between two machine learning models, K-Nearest Neighbours and Random Forest, which were trained with both these labels in order to ascertain whether \textit{AttackType} can produce reliable results in comparison with the \textit{Class} label.


\end{abstract}
\section{Introduction}
In the last few years Intrusion Detection System (IDS) research has had a progressively increasing interest. The application of Artificial Intelligence (AI) methods for IDS has been widely considered in the literature \cite{sultana_survey_2019} due to their ability to learn complex patterns inherent to network-based data. These patterns are later used in the deployment phase to timely detect probable attack attempts that threaten a network's normal functioning and privacy.

Reliable testbeds comprised of both normal network behaviour and attack scenarios are required to train AI algorithms and test their detection performance in controlled environments. The lack of well grounded datasets for the Network Intrusion Detection Systems (NIDS) setting has been appointed as one of the major drawbacks of current research \cite{sharafaldin_toward_2018}. However, in the last few years, some datasets have been proposed to solve this problem, namely the CTU 13 \cite{garcia_empirical_nodate}, the SANTA data set \cite{wheelus_session_2014}, the CICIDS-2017 \cite{sharafaldin_toward_2018}, and the CIDDS-001 dataset \cite{ring_flow-based_nodate}. 

IDS datasets can mainly be divided into two categories, packet-based and flow-based. The more conventional ones such as DARPA99 \cite{thomas_usefulness_2008} or its improved version, KDD CUP 99 \cite{tavallaee_detailed_2009} are packet-based. These contain great amounts of information and features such as packet-headers and payloads \cite{ring_flow-based_nodate}. Flow based datasets, however, such as CTU 13 \cite{garcia_empirical_nodate} and CIDDS-001 \cite{ring_flow-based_nodate} are usually more compact as their features are composed of aggregated information that was extracted from the communications within the network. These types of datasets were proposed by Wheelus and Zuech \textit{et al.} \cite{wheelus_session_2014} in 2014, as they are more recent and there are relativity fewer examples when compared to packet-based ones.

The CIDDS-001 is a recent flow-based dataset for the IDS setting. It contains unidirectional \textit{Netflow} data and was generated using Python scripts to simulate human behaviour on virtual machines of an emulated network. It is very realistic since it respects operating cycles and working hours in enterprises. It contains both normal data and different types of cyber attacks, namely ping scans, port scans, brute forces and  denial of services (DoS). Since the technologies used to generate the attacks are time-dependent the flows of the dataset were labeled based on their timestamp. Four different labels were considered, \textit{Class}, which classifies the flow into normal, attacker, victim, suspicious or unknown, \textit{AttackType}, which represents the type of the executed attack, \textit{AttackID}, which contains the ID of the attack instance and \textit{AttackDescription}, which provides a short description with attack-related details.

In this work, a comparison between the performance of two Machine Learning (ML) models, Random Forest (RF) and K-Nearest Neighbors (KNN), was performed under the CIDDS-001 setting. These models were chosen to conduct this study because they are widely used in the NIDS setting exhibiting great performances in several testbeds \cite{maia_cyber_2021, kumar_development_2020}.

Most studies around this dataset have used the \textit{Class} label as target variable, with one found exception \cite{app11041674}, which used \textit{AttackType}. Therefore, this study seeks to compare both labels and to evaluate if the \textit{AttackType} is a reliable target to train and evaluate ML algorithms in the CIDDS-001 context. 

For a SOC operator, knowing that an attack is occurring is extremely important, but knowing the type of attack is equally important. This knowledge will influence his following decisions as well as what measures to take in order to mitigate the threat. Therefore, a system that is capable of not only detecting attacks but also classifying them is of great use for any IDS.

This paper is organized into multiple sections. Section 2 presents an overview of the current state of the literature regarding NIDS research on the CIDDS-001 dataset. Section 3 describes the algorithms and testbed chosen to perform this study. Section 4 presents the achieved results and their discussion. Section 5 provides the main conclusions of this work as well as future research topics to be consequently addressed.

\section{Related Work}
\begin{sloppypar}
Several anomaly detection approaches based on ML have been proposed in the context of NIDS. Many works have selected the CIDDS-001 as testbed to train and test their algorithms. Most of them used the \textit{Class} label as target variable, presenting outstanding results in model performance. To the best of our knowledge, only one work, \cite{app11041674}, has used the \textit{AttackLabel} label.

In \cite{verma_statistical_2018}, Verma \textit{et al.} performed a statistical analysis of the CIDDS-001 dataset using KNN and K-means clustering. These models were trained using the \textit{Class} label, classifying each flow as either suspicious, attack, victim, unknown or normal. The data used for model training was separated into data from the \textit{External server} and data from the \textit{OpenStack}. The algorithms trained on both sets achieved extremely good results with an accuracy value close to 100\%.

In \cite{8615300}, Althubiti \textit{et al.} trained a Long-Short Term Memory (LSTM) using the CIDDS-001 dataset and the \textit{Class} label, achieving an accuracy of almost 85\% and a recall and precision of almost 86\% and 88\%, respectively. They compared this model with a Support Vector Machine (SVM), a Naïve Bayes and a Multi-Layer Perceptron (MLP) which achieved slightly worse results than the LSTM, around 80\%.

In \cite{verma_machine_2020}, Verma \textit{et al.} tested a variety of ML models to detect DoS attacks in the context of internet of things. The CIDDS-001, UNSW-NB15, and NSL-KDD were used for training and benchmarking with a wide variety of models such as RF, AdaBoost, gradient boosted machine, regression trees and MLPs. The models achieved a good performance in detecting DoS attacks for the CIDDS-001 dataset with the RF presenting an accuracy of almost 95\% and an Area Under the Curve (AUC) of almost 99\%.

In \cite{kilincer_machine_2021}, I. Kilincer \textit{et al.} used five of the most widely acknowledged IDS datasets, CSE-CIC IDS-2018, UNSW-NB15, ISCX-2012, NSL-KDD and CIDDS-001, and trained a  SVM, a KNN and a Decision Tree with each of them. Great results were achieved for every algorithm, with the KNN trained with CIDDS-001 dataset achieving an accuracy, recall, precision and f1-score of approximately 97\%. The model that achieved the best results with this dataset was the Decision Tree with all four referred metrics above 99\%.

In \cite{9133979} Zwane \textit{et al.} performed an analysis of an ensemble learning approach for a flow based IDS using the CIDDS-001 dataset. A variety of ensemble learning approaches were employed. These techniques were applied on three algorithms, Decision Tree, Naïve Bayes and SVMs. A RF was also implemented. Results greatly differed based on the chosen algorithm and ensemble technique. The best performing algorithms were the RF and all the ensembles of decision trees with an accuracy, precision, recall and f1-score of 99\%. The ensembles trained with the Naïve Bayes and the SVMs performed worse, with all four metrics varying between 60\% and 70\%. 

In \cite{adhi_tama_attack_2017}, Tama and Rhee performed an attack classification analysis of IoT network by using deep neural networks for classifying attacks trained using the  CIDDS-001, UNSW-NB15 and GPRS datasets. The deep neural network trained with the CIDDS-001 achieved excelent results, with an accuracy, precision and recall of 99\%, 99\% and 100\% respectively.

In \cite{app11041674}, N. Oliveira \textit{et al.} proposed a sequential approach to intrusion detection by using time-based transformations in the algorithms' input data. The CIDDS-001 dataset was used for training three distinct algorithms, LSTM, RF and MLP. Additionally, the \textit{AttackType} label was chosen instead of the \textit{Class} label. The introduced approach achieved 99\% accuracy for both the LSTM and the RF. An F1-score of near 92\% was obtained for the LSTM.

As far as we are aware, this is the first work to perform a comparison between both labels in an attempt to determine if any gain can be obtained from using \textit{AttackType} instead of the \textit{Class} label.

\end{sloppypar}

\section{Materials and Methods}

This section presents the CIDDS-001 dataset and the way it was labelled, as described by it's authors in the technical report \cite{ring_technical_nodate}. The employed algorithms, RF and KNN, parameters and configurations are described as well as the evaluation metrics chosen for validation and comparison.

\subsection{Dataset Description}

\begin{sloppypar}
The CIDDS-001 (Coburg Intrusion Detection Data Set) was developed by Markus Ring \textit{et al.} \cite{ring_flow-based_nodate} in 2017. This dataset is flow-based unlike most conventional datasets, which are packet-based.

The data present in the CIDDS-001 dataset can mainly be divided into two sets, based on how it was generated. Data generated through \textit{OpenStack} emulating a small business environment and data generated through \textit{External Server},  capturing real network data. The dataset contains a total of 31959175 flows, with 31287934 being generated by the \textit{OpenStack} and 671241 from the \textit{External server}. A total of 33 million flows comprises the CIDDS-001, spanning a total of four weeks of network traffic, from March 3 2017 to April 18 2017. All 33 millions flows are labelled, containing benign network traffic as well as four different types of attacks, Ping Scans, Port Scans, DoS and Brute force. 

Unlike packet-based datasets, which usually contain a lot of features, the CIDDS-001 contains a total of 12 with an additional 4 labels. These are standard \textit{Netflow} features, namely source and destination IPs and ports, transport protocol, duration, date first seen, number of bytes, number of packets, number of flows, type of service and TCP flags. The dataset contains an additional four labels, \textit{Class}, \textit{AttackID}, \textit{AttackType} and \textit{AttackDescription}.

\end{sloppypar}
\subsection{Dataset Labelling}

\begin{figure}[ht]
    \centering
    \includegraphics[width=0.85\textwidth]{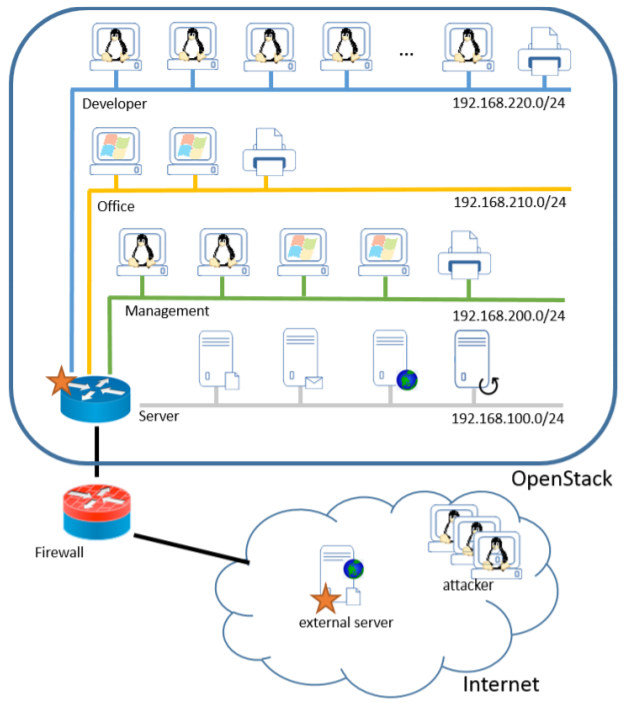}
    \caption{Schematic of the emulated small business environment}{\cite{ring_technical_nodate}}
    \label{fig:cidds-network}
\end{figure}

The labelling process of the CIDDS-001 is composed of two steps, due to the different data sources. Both the traffic from the \textit{External Server} and the one generated by the \textit{OpenStack} environment were processed and labelled accordingly. Since the \textit{OpenStack} traffic was generated in a fully controlled network this involved using the timestamps, origins and targets of the executed attacks \cite{ring_technical_nodate}, marking the remaining traffic as normal. A representation of the simulated small business environment is described in Figure \ref{fig:cidds-network}.

As for the traffic originated from the \textit{External Server}, its labelling process was not as simple. All traffic incoming into this server from the \textit{OpenStack} environment was benign, and as such was labelled as normal. The additional traffic incoming from the three controlled servers, \textit{attacker1, attacker2 and attacker3}, is only malicious, so the traffic incoming and outgoing from these servers is labelled as attacker or victim respectively. All traffic from ports 80 and 443 is labelled as unknown since it comes from a homepage available to anyone interested in visiting. And finally all remaining traffic in the \textit{External server} is labelled as suspicious \cite{ring_technical_nodate}.

The latter labels, unknown and suspicious, due to their uncertain nature, pose a problem when training the models. Considering that this traffic can include both normal and attack entries, and that there is no way to correctly categorize each one, the use of the whole environment as either normal or attack will create a strong bias towards these classes.

\subsection{Dataset Preprocessing and Sampling}

\begin{figure}[ht]
    \centering
    \includegraphics[width=0.85\textwidth]{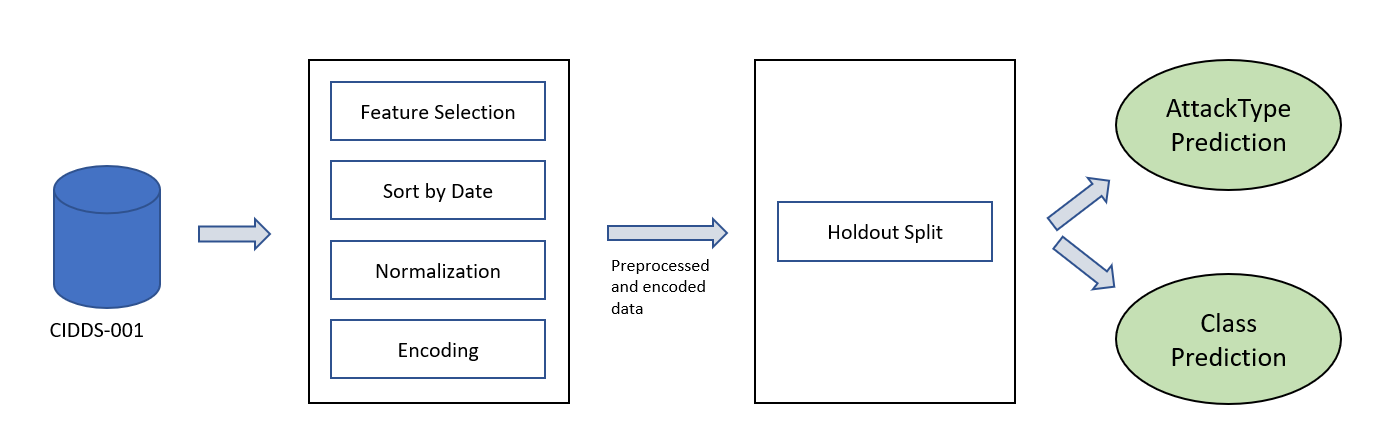}
    \caption{Methodology workflow}
    \label{fig:cidds-worflow}
\end{figure}

In preparation for the ML methods training, the CIDDS-001 has to go through a series of preprocessing procedures. This process was largely based on the work of N. Oliveira \textit{et al.} \cite{app11041674} that implemented a preprocessing pipeline consisting of several steps. A visual representation of the overall process is presented in Figure \ref{fig:cidds-worflow}. An initial analysis to detect errors and inconsistent data was made, eliminating features such as \textit{Flows} that contains a single value throughout the whole dataset and correcting others such as \textit{Bytes} where suffixes like \textit{K} and \textit{M} were replaced by their numerical representation, in order to preserve the sequence of the flow, the dataset was indexed with the \textit{Date first seen} feature. Non-numerical features were encoded using categorical encoding. After applying this pipeline, there were 10 features remaining, \textit{Src IP}, \textit{Src Port}, \textit{Dest IP}, \textit{Dest Port}, \textit{Proto}, \textit{Flags}, \textit{ToS}, \textit{Duration}, \textit{Bytes} and \textit{Packets}, all of them encoded as numbers and scaled between 0 and 1. The sampling process was likewise performed in the same manner as N. Oliveira \textit{et al.} \cite{app11041674}, resulting in a representative sample of the whole dataset. Since the resulting sample is a lot less demanding in terms of memory, processing power and time but still very representative of the dataset as a whole. The sample contains a total of 2,535,456 flows between 2:18:05 p.m., 17 March 2017and 5:42:17 p.m., 20 March 2017. This final sample was then split into a training and testing set using a simple holdout method where 70\% of the data  was used for training and 30\% for testing.

Since this work seeks to compare algorithms trained with both the \textit{AtttackType} and \textit{Class} labels, an additional preprocessing step was performed for the latter. All instances of suspicious, unknown and victim were replaced with the attack class, transforming the problem as a binary classification task with only two categories, normal and attack.

\subsection{Models}

To compare the impact of the \textit{Class} and \textit{AttackType} labels in the ML methods performance, two algorithms were selected to be employed in this work, RF and KNN. These models have been thoroughly studied by the scientific community presenting good results for several IDS datasets \cite{anbar_comparative_2016}.

RF is an ensemble ML method that makes use of several instances of decision trees which are trained with different sub-samples of the same dataset to produce diverse classification rules. When presented with a sample to be classified, all the decision trees of the ensemble make their own prediction. The final classification result is decided by majority voting \cite{biau_random_2016}. The parameters selected for the RF are described in Table \ref{tab:RFParameters}. These values were selected based on grid search within a range of possible values.

\begin{table}[H]
\caption{Random Forest Classifier Parameters.}
\centering
\footnotesize
\begin{tabular}{lr}
\toprule
\textbf{Parameter}	& \textbf{Value}\\
\midrule
Nº of Estimators            & 10\\
Split Criterion             & Gini Impurity\\
Min Samples Split           & 2\\
Min Samples Leaf            & 1\\
Max features                & $\sqrt{n_{features}}$\\
Min Impurity Decrease       & 0\\
Class Weight                & Balanced\\
\bottomrule
\end{tabular}
\label{tab:RFParameters}
\end{table}

On the other hand, KNN is a supervised ML algorithm used for both classification and regression. When used as a classifier, the algorithms prediction is the highest frequency class among the sample's $k$ nearest data points. The distance between the dataset instances is calculated using a specific metric such as Euclidean and Manhattan. The parameters selected for the KNN were also obtained by performing a grid search optimization and are presented in Table \ref{tab:KNNParameters}.

\begin{table}[H]
\caption{KNN Classifier Parameters.}
\centering
\footnotesize
\begin{tabular}{lr}
\toprule
\textbf{Parameter}	& \textbf{Value}\\
\midrule
Nº of Neighbors     & 3\\
Weights             & Uniform\\
Leaf Size           & 30\\
Metric              & Minkowski\\
\bottomrule
\end{tabular}
\label{tab:KNNParameters}
\end{table}

\section{Results and Discussion}

Generally a robust evaluation of ML models considers several metrics \cite{handelman_peering_2019}. In this work four metrics were used: accuracy, precision, recall and F1-score \cite{hossin_review_2015}. Accuracy is a generalist metric that usually gives a good indication of an algorithm's performance. Precision measures the amount of positive predictions that were actually negative. Recall determines how many of the total positive label instances were correctly labelled. F1-score is calculated using both recall and precision, being a good metric to be accounted when a balance between these metrics is intended. Additionally, this metric is extremely reliable when dealing with unbalanced datasets, such as CIDDS-001, since it is not biased towards the majority class \cite{handelman_peering_2019}.

To perform the required computations, Google Colab \cite{colab} was used as hardware support providing free access to considerable amounts of disk space and RAM. However, the hardware is not unlimited and there are no guarantees that it remains the same over time. Therefore, a comparison between the models computing time cost could not be performed.

\subsection{Label Comparison}

The results obtained for the \textit{Class} label were near 100\% for all metrics for both models, RF and KNN. On the other hand, the ones regarding the \textit{AttackType} label are quite distinct being highly influenced by the algorithms performance on the minority classes, such as ping scan and brute force. The macro-averaged results are descibed in Table \ref{tab:AttackTypeGeneralResults}.

\begin{table}[H]
\caption{AttackType Label Macro Average Model Results.}
\centering
\footnotesize
\begin{tabular}{lrrrr}
\toprule
\textbf{Model}  &   \textbf{Accuracy}  & \textbf{Precision}    & \textbf{Recall} & \textbf{F1-Score}\\
\midrule
Random Forest & 0.9560 & 0.9032 & 0.9239 & 0.9134\\
KNN & 0.9694 & 0.9037 & 0.9290 & 0.9161\\
\bottomrule
\end{tabular}
\label{tab:AttackTypeGeneralResults}
\end{table}

Table \ref{tab:AttackTypeRandomForestResults} shows the results of the trained RF model for the classification of each type of attack with the \textit{AttackType} label. 

\begin{table}[H]
\caption{AttackType Label Random Forest Model Results.}
\centering
\footnotesize
\begin{tabular}{lrrrr}
\toprule
\textbf{Class}  &   \textbf{Precision}    & \textbf{Recall} & \textbf{F1-Score}\\
\midrule
Brute Force & 0.9791 & 0.9868 & 0.9239\\
DoS & 1.0000 & 0.9999 & 1.0000\\
No Attack & 0.9999 & 1.0000 & 1.0000\\
Ping Scan & 0.8104 & 0.5344 & 0.6441\\
Port Scan & 0.9906 & 0.9950 & 0.9928\\
\bottomrule
\end{tabular}
\label{tab:AttackTypeRandomForestResults}
\end{table}

Table \ref{tab:AttackTypeKNNResults} displays the same information but for the trained KNN model.

\begin{table}[H]
\caption{AttackType Label KNN Model Results.}
\centering
\footnotesize
\begin{tabular}{lrrrr}
\toprule
\textbf{Class}  &   \textbf{Precision}    & \textbf{Recall} & \textbf{F1-Score}\\
\midrule
Brute Force & 0.9920 & 0.9815 & 0.9867\\
DoS & 0.9999 & 0.9999 & 0.9999\\
No Attack & 0.9999 & 1.0000 & 0.9999\\
Ping Scan & 0.8650 & 0.5406 & 0.6654\\
Port Scan & 0.9900 & 0.9965 & 0.9932\\
\bottomrule
\end{tabular}
\label{tab:AttackTypeKNNResults}
\end{table}


\subsection{Discussion}

As stated in \cite{ring_technical_nodate}, the labelling process of the traffic incoming from the \textit{External Server} in the CIDDS-01 dataset with regards to the \textit{Class} label isn't accurate, as all traffic incoming from ports 80 and 443 is labelled as either unknown or suspicious without any real certainty of whether that flow relates to an attack or not. Additionally a score of nearly 100\% for all metrics in both models trained with this label is abnormal and not coherent, therefore it is highly probable that these models are overfitted and that these results do not reflect their intrusion detection capability.  

As for the results of the \textit{AttackType} label, they appear to be very promising since the labelling process accurately identifies all the attacks that were performed in the testbed. This presents a great advantage over the \textit{Class} label since it assures a more robust and less biased classifier. The obtained results are also lower in terms of macro F1-score since it is typically hard for machine learning algorithms to perform well for the minority classes of unbalanced datasets such as CIDDS-001. Nevertheless, both models performed quite well for all attack types with the exception of the ping scan class.

\section{Conclusion}

This work has established a comparison between two labels, \textit{Class} and \textit{AttackType}, of the CIDDS-001 dataset, a widely regarded testbed for NIDS research. Two ML algorithms, RF and KNN, were trained with both these labels in order to compare their performance as classifiers and assure that the \textit{AttackType} can be a reliable target variable although \textit{Class} is more commonly used in previous works.

The results for the \textit{Class} label were near 100\% for all metrics for both models, seemingly too perfect, suggesting the occurrence of over-fitting. Differently, although the results regarding \textit{AttackType} are slightly lower in terms of absolute values, these seem to be a lot more promising since the labeling process assures the correct identification of all attacks performed in the testbed. The KNN achieved the best F1-score value, 91.61\%, slightly above the one presented by the RF, 91.34\%.

This research, to the best of our knowledge, is the first to address a comparison between both these labels and to appoint that the \textit{Class} labelling process is quite less reliable than the one for the \textit{AttackType}. As future work, the \textit{AttackType} label will be explored in greater detail by experimenting with other ML algorithms in an attempt to improve the presented results.

\section*{Acknowledgements}
This work has received funding from European Union’s H2020 research and innovation programme under SAFECARE Project, grant agreement no.787002.
%
%
%
\bibliographystyle{ieeetr}
\bibliography{bibliography.bib}
%

\end{document}